\documentclass{amsart}
\usepackage{amsmath,amssymb,amscd,graphicx,epsfig}
\usepackage{hyperref}
\usepackage{graphicx}
\usepackage{enumerate}
\usepackage{caption}
\newtheorem{teo}{Theorem}[section]
\newtheorem{lem}[teo]{Lemma}
\newtheorem{cor}[teo]{Corollary}
\newtheorem{prop}[teo]{Proposition}
\newtheorem{remark}[teo]{Remark}

\DeclareMathOperator{\diam}{diam}
\newcommand{\mr}{\mathbb{R}}

\newcommand{\B}{{\mathcal B}}
\newcommand{\Ee}{{\mathcal E}}
\newcommand{\Ii}{{\mathcal I}}
\newcommand{\Ll}{{\mathcal L}}
\newcommand{\Nn}{{\mathcal N}}
\newcommand{\Pp}{{\mathcal P}}
\newcommand{\Ss}{{\mathcal S}}
\newcommand{\Tt}{{\mathcal T}}
\newcommand{\NM}{{\mathbb N}}
\newcommand{\RM}{{\mathbb R}}

\begin{document}
\title{Rapid convergence to  frequency \\for\\ Substitution Tilings of the Plane} 

\pagestyle{headings}
\author{Jos\'e Aliste-Prieto, Daniel Coronel and   Jean-Marc Gambaudo\\ \,\\ \,\ }

\begin{abstract} This paper concerns self-similar tilings of the Euclidean plane. We consider the number of occurrences of a given tile in any domain bounded by a Jordan curve. For a large class of self-similar tilings, including many well-known examples,  we give estimates of the oscillation of this number of occurrences around its average frequency times the total number of tiles in the domain, which depend only on the Jordan curve. 
\end{abstract}

\noindent \address{ {\it J. Aliste-Prieto: }
Centro de Modelamiento Matem\'atico, Universidad de Chile, Blanco Encalada 2120 7to. piso,
Santiago, Chile.}
\email{jaliste@dim.uchile.cl}
\noindent \address{ {\it D. Coronel: } 
Facultad de Matem\'aticas, Pontificia Universidad Cat\'olica de Chile, Campus San Joaqu\'in, Avenida Vicu\~{n}a Mackenna 4860, 
Santiago, Chile.}
\email{acoronel@mat.puc.cl}

\noindent \address{ {\it J.-M. Gambaudo: } Laboratoire j.-A.
   Dieudonn\'e, Universit\'e de Nice - Sophia Antipolis-CNRS, 06108
   Nice Cedex 02, France.}  \email{gambaudo@unice.fr}

\date{\today}
\maketitle
\markboth{J. Aliste-Prieto,  D. Coronel, J.-M. Gambaudo}{Speed of convergence for self-similar tilings}

\section{Introduction}
Quasicrystals are alloys that show long range order but possess symmetries that 
prevent them from being crystals. From their spectacular experimental realization
in  the early 80's \cite{shechtman:1984} to their very recent discovery as natural
 objects in the Kamtchatka mountains \cite{bindi:2009}, quasicrystals have 
been the subject of very active research, whose application and interest extend
far beyond the scope of solid state physics. The first examples observed were rapidly
quenched alloys of Aluminum and Manganese exhibiting icosahedral symmetry. In these
 quasicrystals, atoms are known to appear with a given frequency. This means that for 
a large ball $B$ of radius $R$, the ratio of the number $N_s$ of atoms inside $B$
corresponding to a specific atomic element $s$ (for example  Aluminum in the first known cases)
to the total number $N$ of atoms in the ball $B$ tends to a limit $\nu_s$ as $R$ tends to infinity:

\[\lim_{R\to {+\infty}} \frac {N_s}{N} \,=\, \nu_s.\]
 
For such examples, the following questions arise naturally:
\begin{enumerate}
\item to estimate the oscillation of the ratio ${N_s}/{N}$ around its limit $\nu_s$, that is to say, to give upper bounds for the speed of convergence to $0$ of the ratio $\vert  {N_s}/{N}\, -\, \nu_s\vert$ as $R$ goes to infinity; 
 \item to extend these estimates to domains whose closure are not Euclidean balls.
 \end{enumerate}
We now make these questions more concrete by using aperiodic tilings of the Euclidean space to model quasicrystals (see \cite{Bellissard}), for details and precise statements of the results, see Section \ref{sec:definitions}. Consider a tiling  $\Tt$ of $d$-dimensional Euclidean space $\mr^d$ made with isometric copies of a finite set of tiles
$\{p_1, \dots, p_n\}$, where the $p_i$'s are homeomorphic to a closed ball in $\mr^d$. Let $V$ be a (large) set  in $\mr^d$ also homeomorphic to a closed ball and let $\partial V$ be its boundary. For each $i \in \{1,\ldots , n\}$, we let $\Nn(V, \Tt, p_ i)$ denote the number of isometric copies of $p_i$ in $V$, $\Nn(V, \Tt)$ the number of tiles of $\Tt$,  and 
 $\Ll(\partial V, \Tt)$ the number of tiles of $\Tt$ that intersect $\partial V$.
If $V$ is a ball, then it is well-known that there are tilings such that for each 
$i$ in $\{1, \dots , n\}$, the ratio $\Nn(V, \Tt, p_ i)/\Nn(V, \Tt)$ tends to a well-defined 
limit $\nu_i$  as the radius of $V$ tends to infinity (see for instance \cite{GS,Sol,LP}). 
In fact, this is usually a consequence of the unique ergodicity of the associated dynamical system 
(see for instance \cite{LMS}). However, in general, very little is known concerning an upper bound for the quantity
$\vert\Nn(V, \Tt, p_ i)\, -\, \nu_i.\Nn(V, \Tt)\vert$, specially when $V$ is not a Euclidean ball 
(see \cite{LP,AlisteCoronel} for the case when $V$ equals a ball or box). In particular, for an aperiodic tiling, can we hope to obtain an estimate as strong as the one we can formulate for periodic tilings which reads:
\[ (*)\quad \vert\Nn(V, \Tt, p_ i)\, -\, \nu_i\Nn(V, \Tt)\vert\, \leq K\Ll(\partial V, \Tt)\]
for some $K>0$?

In this paper, we answer this question  in the affirmative for a large class of self-similar tilings 
in two dimensions. Self-similar tilings, which are probably the most studied examples of aperiodic 
tilings (see for instance \cite{Radin,Sol,AP,Rob,Cs} and references therein, examples are given in 
Section \ref{examples}), are associated with a (substitution) primitive  square matrix $M$  
having non-negative integer elements. 
By Perron-Frobenius theory (see for instance \cite{Seneta, HornJohnson}, see also 
Subsection \ref{sec:PF} for more details), we know that there exists a largest positive real eigenvalue $\mu$, the remainder of the spectrum being in a ball centered at $0$ with radius smaller than $\mu$. We denote by $r(M)$  the modulus of the second largest eigenvalue of $M$,  that is, 
\[r(M) = \max \{|\eta|\mid \eta \neq \mu\text{ is an eigenvalue of } M \}.\]
The simple condition
$r(M) < \sqrt{\mu}$  ensures that the above estimate $(*)$ holds true for the associated 
self-similar tiling. If  $r(M) =\sqrt{\mu}$  and if all eigenvalues with modulus $r(M)$ are semi-simple
(their algebraic and geometric multiplicities coincide, 
see for instance \cite{IJ} for more details), then the estimate $(*)$ has to be relaxed to:
\[ (**)\quad \vert\Nn(V, \Tt, p_ i)\, -\, \nu_i\Nn(V, \Tt)\vert\, \leq K\Ll(\partial V, \Tt)\ln{\Ll(\partial V, \Tt)}\]
for some $K>0$. Some explicit examples of self-similar 2-dimensional tilings satisfying all these conditions will be given at the end of the paper.

These results are more explicitly stated in the next section. Ideas for the proofs have been inspired by   J. Peyri\`ere's 1986 paper \cite{Peyriere} and can be related to B. Adamcezswki's work \cite{boris}. 

\section{Definitions and Main result}
\label{sec:definitions}
\subsection{Definitions}
Let $\Lambda$ be a closed subset of the Euclidean plane $\RM^2$.  
A \emph{tiling} of  $\Lambda$ is a countable collection $\Tt =(t_j)_{j\in J}$ 
of closed subsets of $\Lambda$ that cover $\Lambda$ and have pairwise disjoint interiors. The sets 
$t_j$ are called \emph{tiles} and, in this paper, all tiles are supposed to be homeomorphic to the closed unit ball in $\mr^2$, see Section \ref{examples} for examples.

Let $\Ee$ be a group of isometries on the plane that contains the group of translations. A tile $q$ is \emph{$\Ee$-equivalent} to a tile $p$ (or is a \emph{$\Ee$-copy} of $p$) if $q$ is the image of $p$ by an isometry in $\Ee$. Let $\Pp =\{p_1, \dots, p_n\}$ be a finite collection of tiles. A tiling $\Tt$ of  $\Lambda \subseteq \mr^2$ is 
\emph{$\Ee$-generated} by $\Pp$ if every tile in $\Tt$ is $\Ee$-equivalent to some tile in $\Pp$. The set of all tilings of $\Lambda$ that are $\Ee$-generated by $\Pp$ is denoted by $\Omega_{\Ee,\Pp}(\Lambda)$. When $\Lambda = \mr^2$, we write $\Omega_{\Ee,\Pp}$ instead of $\Omega_{\Ee,\Pp}(\mr^2)$.

Let $\lambda > 1$. Given a subset $U\subseteq \mr^2$, let $\lambda U:=\{\lambda x \mid x\in U\}$.
A \emph{substitution rule} (with dilation factor $\lambda$) is a collection $\Ss = (\Ss_i)_{i=1}^n$, where $\Ss_i$ belongs to 
$\Omega_{\Ee,\Pp}(\lambda p_i)$ for all $i\in\{1,\ldots,n\}$. Let $\Ss$ be a substitution rule. On one hand, the dilation by $\lambda$ induces a natural map from $\Omega_{\Ee, \Pp}(\Lambda)$ to 
$\Omega_{\Ee, \lambda\Pp}(\lambda\Lambda)$, where $\lambda\Pp = \{\lambda p \mid p\in\Pp\}$. On the other hand, $\Ss$ induces a map from 
$\Omega_{\Ee, \lambda\Pp}(\lambda\Lambda)$ to $\Omega_{\Ee, \Pp}(\lambda\Lambda)$, which is defined by subdividing the tiles of a tiling in 
$\Omega_{\Ee, \lambda\Pp}(\lambda\Lambda)$ according to the substitution rule.  
For details see Subsection \ref{sec:substitution}. The composition of these two maps yields a map 
$\Ii_{\Ss,\Lambda}: \Omega_{\Ee, \Pp}(\Lambda)\to \Omega_{\Ee, \Pp}(\lambda\Lambda)$. If $\Lambda =\mr^2$, then $\Ii_\Ss:=\Ii_{\Ss,\mr^2}$ is a 
self-map of $\Omega_{\Ee, \Pp}$ and is referred to as the \emph{substitution map}. 
A tiling $\Tt$ is called \emph{self-similar} if it is a periodic point for $I_{\Ss}$. 
A tiling $\Tt$ in $\Omega_{\Ee,\Pp}$ is \emph{admissible} for $\Ss$ if it belongs to
\[\Omega_{\Ss} := \bigcap_{k\geq 0}\Ii_{\Ss}^k(\Omega_{\Ee, \Pp}).\]

The \emph{substitution matrix} associated with $\Ss$ is the $n$-by-$n$ matrix $M_{\Ss}=(m_{i,j})_{i,j}$, where for each $i,j\in\{1,\ldots,n\}$, the coefficient $m_{i,j}$ is the number of $\Ee$-copies of $p_i$ in $\Ss_j$. Recall that a matrix $M$ is \emph{primitive} if there exists $n>0$ such that all the elements of $M^n$ are positive. The substitution rule $\Ss$ is \emph{primitive} if its substitution matrix is primitive. In this paper, all the substitution rules considered will be primitive. 

\subsection{Main result}
Let $\Pp=\{p_1,\ldots,p_n\}$ be a finite set of tiles, $\Ss$ a substitution rule with dilation factor $\lambda>1$
and $\Tt$ an admissible tiling for $\Ss$.  Given a Jordan curve $\Gamma$ in $\mr^2$ 
bounding a topological closed disk $\Delta$, 
we let $\Ll(\Gamma, \Tt)$ denote the number of tiles of $\Tt$ that intersect $\Gamma$,   
$\Nn(\Delta, \Tt)$ be the number of tiles of $\Tt$ included in $\Delta$, and for each $i \in \{1,\ldots , n\}$,
we define $\Nn(\Delta, \Tt, p_ i)$ 
to be the number of $\Ee$-copies of $p_i$ contained in $\Delta$. 

The following theorem, which constitutes the main result of this paper, provides   estimates on the oscillation of the number of occurrences $\Nn(\Delta, \Tt, p_i)$
around an average frequency, which only depends on the Jordan curve $\Gamma$ bounding $\Delta$.

\begin{teo}\label{main}
Let $\Pp =\{p_1, \dots, p_n\}$ be a finite collection of tiles, $\Ss$ a substitution rule 
with dilation factor $\lambda>1$ and primitive substitution matrix $M_\Ss$. 
There exist positive numbers $\nu_1,\ldots , \nu_n $ depending only on 
the substitution matrix $M_\Ss$ that satisfy: 
\begin{itemize}
\item [(i)] If $r(M_{\Ss}) < \lambda$, then there exists $K >0$  such that, 
for every Jordan curve $\Gamma$ bounding a topological closed disk $\Delta$  and every tiling $\Tt$ in  
$\Omega_\Ss$, we have:
\[\vert \Nn(\Delta, \Tt,  p_ i) \, -\, \nu_i \Nn(\Delta ,\Tt)\vert\, \leq\, K \Ll(\Gamma,\Tt).\]

\item[(ii)] If $r(M_{\Ss}) = \lambda$ and all the eigenvalues with 
modulus $\lambda$ are semi-simple (that is, their algebraic and geometric multiplicities coincide), 
then there exists $K>0$  such that, 
for every Jordan curve $\Gamma$ bounding a topological closed disk $\Delta$  and every tiling $\Tt$ in  
$\Omega_\Ss$, we have:
\[\vert \Nn(\Delta,\Tt,  p_i) \, -\, \nu_i\Nn(\Delta,\Tt)\vert\, \leq\, 
K \Ll(\Gamma,\Tt) \ln\Ll(\Gamma,\Tt)).\]
\end{itemize}
\end{teo}

\begin{remark}
Notice that Theorem \ref{main} is given for a quite general setting: 
 \begin{enumerate}
 \item On one hand, we do not require the tiles in $\Pp$ to be polygons, 
they need not even be disks with piece-wise smooth boundaries. 
In fact, the boundary of the tiles could be Jordan curves with infinite 
length and positive Hausdorff dimension, as is the case for Rauzy tilings 
that we shall discuss later.
 
 \item   On the other hand, we do not require the standard finite pattern  
condition used in tiling theory (see \cite{KP}). 
This allows us to deal with self-similar tilings  
which have ``fault lines'' along which tiles can slide past one another. 
This was the case for the self-affine tilings studied by N. Priebe Frank and L. Sadun \cite{PSa}.  
\end{enumerate}
\end{remark}
\begin{remark}
For a given substitution, the estimates given by Theorem \ref{main} may depend on the group $\Ee$ being considered. This is the 
case of Penrose tilings, see Section \ref{examples} for details.
\end{remark}
\begin{remark} \label{bla}Theorem \ref{main} can be easily extended to colored tilings (for an example, see the square tilings in Section \ref{examples}). 
\end{remark}

\section{The tools}
\subsection{Jordan curves and locally finite tilings}
In this section, we provide some simple combinatorial estimates for the number of tiles of a tiling that intersect a Jordan curve.  
First we must define some notation. Denote by $B_x(r)$ the closed ball of radius $r$ around $x$ in $\mr^2$. 
Given a subset $U\subseteq\mr^2$, define
\[r_\text{min}(U) := \sup\{r>0 \mid \text{there exists } x\in U \text{ s.t. }B_x(r)\subseteq U\}.\]

To  a tiling $\Tt$ of $\RM^2$, we associate 
\[r_\Tt=\inf\{r_\text{min}(t)\mid t\in\Tt\}\]  
and 
\[R_{\Tt}=\sup\{\diam(t)\mid{t\in\Tt}\}/2.\]

The following lemma shows that if $r_\Tt$ is positive and $R_\Tt$ is finite, then there is a uniform bound 
on the number of tiles intersected by balls of a prescribed radius. 
\begin{lem}
\label{lem:finite_complexity}
Suppose that  $\Tt$ is a tiling with $R_\Tt<+\infty$ and $r_\Tt>0$. Then every ball of radius $2R_\Tt$ intersects at most 
$K_\Tt:=\lfloor 16R_\Tt^2 r_\Tt^{-2}\rfloor$ tiles of $\Tt$, where $\lfloor -\rfloor $ stands for the integer part.
\end{lem}
\begin{proof} Fix $x\in\mr^2$. It is clear that each tile of $\Tt$ intersecting $B_x(2R_\Tt)$ is included 
in $B_x(4R_\Tt)$. Since each one of these tiles contains a ball of radius $r_\Tt$, 
comparing area yields $r_\Tt^2\pi N\leq  16\pi R_\Tt^2$, where $N$ is the number of tiles intersecting $B_x(2R_\Tt)$ and 
the conclusion follows. 
\end{proof}

By virtue of the previous lemma, we say that a tiling $\Tt$ is \emph{locally finite} if $R_\Tt<+\infty$ and $r_\Tt>0$. The next result provides an estimate for the diameter of a simple curve in terms of the number of tiles of a locally finite tiling that the curve intersects.

To simplify notation, it is convenient to identify curves with their images. Let $\gamma:[0,1]\rightarrow\mr^2$ be a simple curve. The identification induces an order on the image: Let $y,y'$ in $\gamma$ with $y = \gamma(s)$ and $y' = \gamma(t)$.  If $\gamma$ is open, then $y\leq y'$ if and only if  $0\leq s\leq t \leq 1$; if $\gamma$ is closed, then $y \leq y'$ if and only if $0\leq s\leq t<1$. 
 \begin{lem}
\label{lem:diam_jordan_arc}
Let $\Tt$ be a locally finite tiling  and $\gamma$ be a simple curve.  
Then, 
\[ \operatorname{diam}(\gamma)\leq 2R_\Tt \Ll(\gamma,\Tt),\]
where  $\Ll(\gamma,\Tt)$ denotes the number of tiles that $\gamma$ intersects.
\end{lem}
\begin{proof}
By compactness, 
there exist $x, x'\in\gamma$ with $x\leq x'$ and $d(x, x') = \diam(\gamma)$. 
We construct a sequence of points $(x_i)_{i=0}^{\ell}$ as follows. 
Fix $\varepsilon>0$ and set $x_0 = x$. To construct $x_1$, if $x'$ belongs to 
$B_{x_0}(2R_\Tt+\varepsilon)$, then set $\ell = 1$ and $x_1 = x'$. 
If $x'$ does not belong to $B_{x_0}(2R_\Tt+\varepsilon)$, then define $x_1$ by 
\[x_1 := \max \{z\in\gamma \mid z \in B_{x_0}(2R_\Tt+\varepsilon)\}.\]
It is clear that $d(x_1, x_0) = 2R_\Tt+\varepsilon$. This construction can be extended inductively to 
obtain a sequence $x = x_0<x_1<\ldots<x_{\ell-1}<x_\ell = x'$ in $\gamma$ such that  
for all $k\in \{1, \dots, \ell-1\}$:
\begin{itemize}
\item  $d(x_{k-1}, x_{k}) = 2 R_\Tt + \varepsilon$,
\item $\{z\in\gamma\mid x_{k}\leq z\leq {x'}\} \cap \cup_{m=0}^{k-1} B_{x_m}(2R_\Tt+\varepsilon) = \{x_{k}\},$
\end{itemize}
and such that $d(x_{\ell-1}, x_\ell)\leq 2 R_\Tt+\varepsilon$ (see Figure \ref{fig:diam_jordan_arc}).

For each $k\in\{0, \dots, \ell-1\}$, choose a tile $t_{j(k)}$ in $\Tt$ that contains $x_k$. 
Since $d(x_{k},x_{k'})\geq 2R_\Tt+\varepsilon$ for all $k,k'$ in $\{0,\ldots,\ell-1\}$, 
it follows that $t_{j(k)}$ is not the same tile as $t_{j(k')}$ unless $k=k'$. 
Hence, the number $\Ll(\gamma, \Tt)$ of tiles in $\Tt$ that intersect 
$\gamma$ is at least $\ell$ (see  Figure \ref{fig:diam_jordan_arc}).
Moreover, it is easy to check that $\ell(2R_\Tt +\varepsilon) \geq d(x,x') = \diam(\gamma)$. 
Thus, using the estimate for $\Ll(\gamma,\Tt)$ we get
  \[ \diam (\gamma) \leq  (2R_\Tt+\varepsilon)\Ll(\gamma, \Tt),\]
  and the conclusion can be obtained by letting $\varepsilon$ go to $0$.
\end{proof}
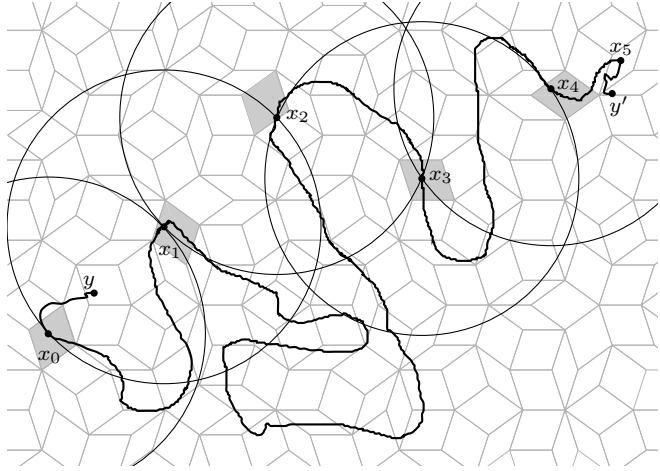
\begin{figure}
\begin{center}
\begin{picture}(0,0)%
\includegraphics[scale=0.50]{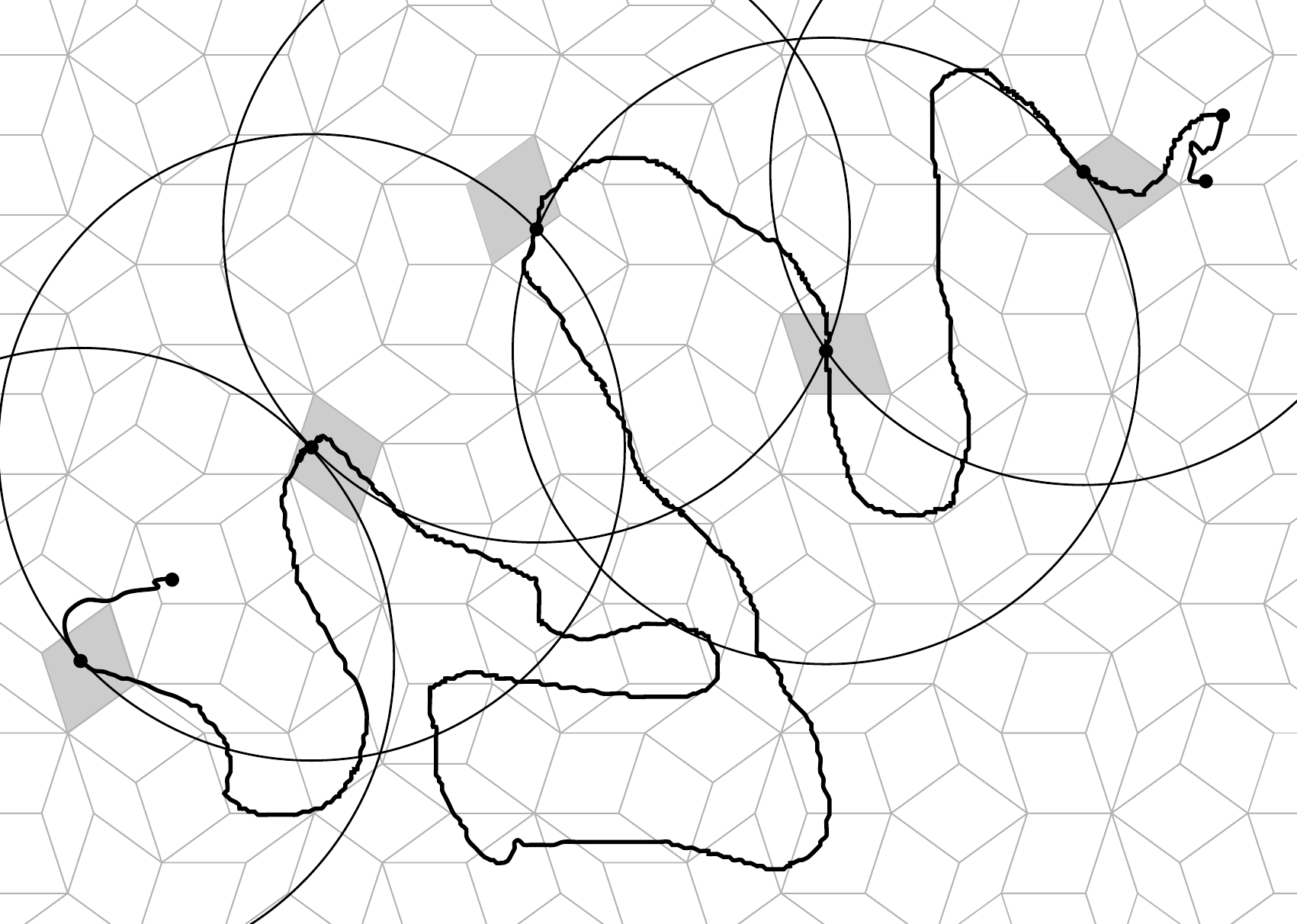}
\end{picture}%
\setlength{\unitlength}{1cm}%
\begingroup\makeatletter\ifx\SetFigFontNFSS\undefined%
\gdef\SetFigFontNFSS#1#2#3#4#5{%
  \reset@font\fontsize{#1}{#2pt}%
  \fontfamily{#3}\fontseries{#4}\fontshape{#5}%
  \selectfont}%
\fi\endgroup%
\begin{picture}(10,6)(0,0)
\put(1,2.4){\makebox(0,0)[lb]{\smash{{\SetFigFontNFSS{8}{9.6}{\rmdefault}{\mddefault}{\updefault}{$y$}}}}}
\put(0.4,1.4){\makebox(0,0)[lb]{\smash{{\SetFigFontNFSS{8}{9.6}{\rmdefault}{\mddefault}{\updefault}{$x_0$}}}}}
\put(2,2.8){\makebox(0,0)[lb]{\smash{{\SetFigFontNFSS{8}{9.6}{\rmdefault}{\mddefault}{\updefault}{$x_1$}}}}}
\put(3.7,4.6){\makebox(0,0)[lb]{\smash{{\SetFigFontNFSS{8}{9.6}{\rmdefault}{\mddefault}{\updefault}{$x_2$}}}}}
\put(5.6,3.8){\makebox(0,0)[lb]{\smash{{\SetFigFontNFSS{8}{9.6}{\rmdefault}{\mddefault}{\updefault}{$x_3$}}}}}
\put(7.3,5){\makebox(0,0)[lb]{\smash{{\SetFigFontNFSS{8}{9.6}{\rmdefault}{\mddefault}{\updefault}{$x_4$}}}}}
\put(8,5.5){\makebox(0,0)[lb]{\smash{{\SetFigFontNFSS{8}{9.6}{\rmdefault}{\mddefault}{\updefault}{$x_5$}}}}}
\put(8,4.7){\makebox(0,0)[lb]{\smash{{\SetFigFontNFSS{8}{9.6}{\rmdefault}{\mddefault}{\updefault}{$y'$}}}}}
\end{picture}%
\caption{Proof of Lemma \ref{lem:diam_jordan_arc}: Construction of the sequence 
$x_1,\ldots,x_l$ and $t_{j(1)},\ldots,t_{j(l)}$.}
\label{fig:diam_jordan_arc}
\end{center}
\end{figure}
Given a Jordan curve $\Gamma$ and two tilings $\Tt$ and $\Tt'$, the next lemma compares the number of tiles of 
$\Tt$ that $\Gamma$ intersects with the number of tiles of $\Tt'$ that $\Gamma$ intersects. 
\begin{lem}
\label{lem:bord}
Let $\Gamma$ be a Jordan curve and $\Tt$ be a locally finite tiling. Then, for every locally finite tiling 
$\Tt'$ such that  $R_{\Tt'} > R_\Tt$  and $K_{\Tt'}<\Ll(\Gamma,\Tt')\leq \Ll(\Gamma,\Tt)$, we have:
\[\Ll(\Gamma,\Tt') \leq (2K_{\Tt'} + 1) \frac{R_\Tt}{R_{\Tt'}} \Ll(\Gamma,\Tt),\]
where $K_{\Tt'}$ is the constant defined in Lemma \ref{lem:finite_complexity}. 
\end{lem}
\begin{proof}
For every point $y\in\mr^2$, we define 
\[C_{y} := \{t\in\Tt'\mid t\cap B_{2R_\Tt'}(y)\neq\emptyset\}\quad\text{and}\quad\widehat{C}_y:= \cup_{t\in C_y} t.\]
We construct a collection $\mathcal{Y}=\{y_i\}_{i=1}^{p}$ (with $p$ to be determined) of 
points in $\Gamma$ such that the collection $\{\widehat{C}_{y_i}\}_{i=1}^p$ covers $\Gamma$. Fix any point of $\Gamma$ as $y_1$. Now suppose that $y_1,\ldots,y_j$ 
have been chosen such that, for all $i\in\{1,\ldots,j\}$ and $k\in\{1,\ldots,i-1\}$, the point 
$y_i$ does not belong to $\widehat{C}_{y_k}$. There are two cases to consider. 
Either the sets $\{\widehat{C}_{y_i}\}_{i=0}^{j}$ cover $\Gamma$, in which case we set $p=j$ and the construction is completed; or they do not cover $\Gamma$, in which case we choose any point of 
$\Gamma\setminus\cup_{i=0}^{j} \cup_{t\in C_{y_i}} t$ to be $y_{j+1}$  and continue iterating the construction. 
Observe that $\Ll(\Gamma,\Tt')$ is finite because $\Tt'$ is locally finite and $\Gamma$ is compact, and in each iteration, we add at least one tile of $\Tt'$ that intersects $\Gamma$ to the area covered by $\{\widehat{C}_{y_i}\}_{i=0}^{j}$. Hence, the construction stops in finitely many steps. 

On one hand, by Lemma \ref{lem:finite_complexity}, each $C_{y_i}$ contains at most 
$K_{\Tt'}$ tiles of $\Tt'$. It follows that 
\begin{equation}
\label{eq:est1_p}
\Ll(\Gamma,\Tt')\leq p K_{\Tt'},
\end{equation}
and since $K_{\Tt'}<\Ll(\Gamma,\Tt')$, we deduce that $p\geq 2$.
 
On the other hand, fix $i\in\{1,\ldots,p\}$.  Hence, there exists  
$j\in\{1,\ldots,p\}$ such that $y_j$, which belongs to $\Gamma$, does not belong to 
$B_{y_{i}}(R_{\Tt'}-R_\Tt)$. Suppose that $y_i < y_j$ (the other case is analogous), 
then there is a point $y_i'>y_i$ in $\Gamma$ with $d(y_i, y_i') = R_{\Tt'}-R_{\Tt}$ 
such that the arc $[y_i,y_i']_\Gamma=\{z\in\Gamma \mid y_i \leq z \leq y_i'\}$, which joins $y_i$ and $y_i'$, is contained by the ball $B_{y_{i}}(R_{\Tt'}-R_\Tt)$. Since $i$ was arbitrary, using Lemma \ref{lem:diam_jordan_arc} yields 
\[ R_{\Tt'}-R_\Tt\leq\diam([y_i,y_i']_\Gamma)\leq 2R_\Tt  \Ll([y_i,y_i']_\Gamma, \Tt)\]
for all $i\in\{1,\ldots,p\}$. Combining all these inequalities, we obtain
\begin{equation}
\label{eq:est2_p0}
p (R_{\Tt'}-R_\Tt) \leq 2R_\Tt \sum_{i=1}^p \Ll([y_i,y_i']_\Gamma, \Tt).
\end{equation}
From the construction of $\mathcal{Y}$, it is clear that $d(y_i, y_j) > 2R_{\Tt'}$ for all $i<j$, and, in particular,  the collection $\mathcal{B}=\{B_{y_i}(R_{\Tt'}-R_\Tt)\}_{i=1}^{p}$ is pairwise disjoint. Moreover, the distance between two different balls in $\B$ is greater than $2R_\Tt$. It follows that no tile of $\Tt$ may intersect more than one ball in $\B$. Hence, no tile of $\Tt$ may intersect more than one arc $[y_i,y_i']_\Gamma$. Thus, combining  \eqref{eq:est2_p0} and \eqref{eq:est1_p}  we get 
\begin{equation}
\label{eq:est2_p}
\Ll(\Gamma,\Tt') \leq 2K_{\Tt'}\frac{R_\Tt}{R_{\Tt'}-R_\Tt}   \Ll(\Gamma, \Tt).
\end{equation}
To finish the proof, fix $c>0$ arbitrarily and consider the following two cases. First, suppose that $R_{\Tt'}/R_{\Tt} \leq 1 + c$. Since $\Ll(\Gamma,\Tt')\leq \Ll(\Gamma,\Tt)$, it follows that 
\begin{equation}
\label{eq:menorquec}
\Ll(\Gamma,\Tt')\leq (1+c)\frac{R_\Tt}{R_{\Tt'}}\Ll(\Gamma,\Tt).
\end{equation}
Now suppose that $R_{\Tt'}/R_{\Tt} > 1 + c$. It is not difficult to check that 
\[(1+c)(R_{\Tt'} - R_{\Tt}) > c R_{\Tt'}.\] 
Replacing this inequality in  
\eqref{eq:est2_p} we get 
\begin{equation}
\label{eq:est2_p2}
\Ll(\Gamma,\Tt') \leq 2K_{\Tt'} \frac{R_\Tt}{R_{\Tt'}}\left(1+\frac{1}{c}\right)\Ll(\Gamma,\Tt).
\end{equation}
Combining \eqref{eq:est2_p2} and \eqref{eq:menorquec} yields
\[\Ll(\Gamma,\Tt')\leq \max\left\{1+c,2K_{\Tt'}\left(1+\frac{1}{c}\right)\right\}\frac{R_\Tt}{R_{\Tt'}}\Ll(\Gamma,\Tt).\]
An easy computation shows that the last bound is optimal when $c = 2K_{\Tt'}$ and the conclusion follows.
\end{proof}
\subsection{Hierarchical sequences}
\label{sec:substitution}
In this subsection, we recall the concept of ``hirarchical sequence'' of a tiling. 
This notion have been used in the context of  self-similar tilings satifying the 
standard finite pattern condition  by many authors, see for 
instance \cite{KP}. Here, we extend its use to admissible tilings.

Let $\Ss$ be a substitution rule with dilation factor $\lambda$. For each $l\in\NM$, 
let $\lambda^l\Pp=\{\lambda^l p \mid p\in \Pp\}$ and consider the map 
$D_l: \Omega_{\Ee,\lambda^{l-1}\Pp} \rightarrow \Omega_{\Ee,\lambda^{l}\Pp}$ defined by 
$D_l(\Tt) = \lambda\Tt$. Also consider the \emph{decomposition} maps 
$S_l: \Omega_{\Ee,\lambda^{l}\Pp}\rightarrow \Omega_{\Ee,\lambda^{l-1}\Pp}$ defined by 
\[S_l (\Tt) = \{ g(\lambda^{l-1}\Ss_i)\mid g\in \Ee, i\in\{1,\ldots,n\}, g(\lambda^l p_i)\in\Tt\}\]
for each $l\in\NM^*$. By definition, $\Ii_\Ss  = S_1 \circ D_1$. 

It is not difficult to check that $\Ii_\Ss$ is onto when restricted to $\Omega_{\Ss}$. 
This implies that for each admissible tiling $\Tt$, there is a sequence $(\Tt^l)_{l\in\NM}$ of tilings,
called a \emph{hierarchical sequence} of $\Tt$, such that $\Tt^0 = \Tt$ and $\Tt^{l-1} = S_l(\Tt^{l})$ 
for all $l\in\NM$, that is, each tile of $\Tt^{l}$ can be decomposed into tiles of $\Tt^{l-1}$ 
according to the substitution rule $\lambda^{l-1}\Ss$. This sequence is constructed as follows: 
Set $\Tt^0 = \Tt$ and, for each $l>0$, inductively choose 
$\bar{\Tt}^l \in I_\Ss^{-1}(\bar{\Tt}^{l-1})$ and then set 
$\Tt^l = \lambda^l\bar{\Tt}^l$. It is not difficult to check that 
$\lambda^{l-1}S_1(\lambda^{1-l}\Tt') = S_l(\Tt')$ for every $\Tt'$ in $\Omega_{\Ee,\lambda^l\Pp}$. 
It follows that $\Tt^{l-1} = S_l(\Tt^l)$ for all $l\in\NM$.

\begin{remark}Notice that all tilings in $\Omega_{\Ss}$ are locally finite and that for each tiling $\Tt$ and each $l > 0$, $\Tt^l$ is also locally finite and satisfies:
\[r_{\Tt^l} = \lambda^l r_\Tt, \quad R_{\Tt^l} = \lambda^l R_\Tt, \text{and thus}\quad  K_{\Tt^l} = K_{\Tt}.\]
\end{remark}

\begin{prop} \label{prop.hier}
Let $\Tt$ be an admissible tiling for $\Ss$. For every topological closed disk $\Delta$ there is a finite collection $\Delta_0,\dots , \Delta_{m-1}$ of closed subsets of $\Delta$ such that:
\begin{enumerate}[(i)]
\item $\cup_{l=0}^{l=m-1}\Delta_i  =\cup_{j\in J_\Delta}t_j$ where $J_\Delta =\{j \, \vert\,  t_j\subset \Delta\}$. 
\item   All $\Delta_l$'s  have pairwise  disjoint interiors.
\item For each $l\in \{0, \dots, m-1\}$, $\Delta_l$ is a union of  tiles in $\Tt^l$, which does not contain a tile in $\Tt^{l+1}$.
\item $\Delta$  does not contain a tile of $\Tt^m$. 
\end{enumerate}
The collection  $\Delta_0,\dots , \Delta_{m-1}$ is called a {\it hierarchical decomposition} of the closed disk  $\Delta$.
Moreover, if $\Gamma$ is the Jordan curve bounding $\Delta$, then 
\begin{equation}
\label{lem:lambda_m}
\lambda^{m - l - 1}\leq \frac{R_\Tt}{r_\Tt} \Ll(\Gamma,\Tt^l)\quad \text{and} \quad \Nn(\Delta_l,\Tt^l) \leq\, \|M\|_1\Ll(\Gamma,\Tt^{l+1})
\end{equation}
for all $l\in\{0,\ldots,m-1\}$, where $\|M\|_1$ is the maximum absolute column sum of the substitution rule $M$. 
\end{prop}
\noindent 
\begin{proof}
Choose $m$ to be the smallest integer such that no tile in $\Tt^m$  is included in $\Delta$ and set 
\[P_{m-1} = \{ t\in \Tt^{m-1}\mid t\subseteq \Delta\}\quad\text{and} \quad \widehat{\Delta}_{m-1} = \cup_{t\in P_{m-1}} t.\]
Applying  the recursion 
\[P_{l-1} = \{t\in \Tt^{l-1}\mid t\subseteq \overline{\Delta\setminus\widehat{\Delta}_l}\}\quad \text{and}\quad \widehat{\Delta}_{l-1} = \widehat{\Delta}_{l} \cup (\cup_{t\in P_{l-1}} t),\]
for $l\in\{1,\ldots,m-1\}$, we obtain a sequence of sets $P_0,\ldots,P_{m-1}$ and $\widehat{\Delta}_0,\ldots,\widehat{\Delta}_{m-1}$, and it is straight forward to check that the sequence 
$\Delta_l = \cup_{t\in P_{l}} t$ for all $l\in\{0,\ldots,m-1\}$ satisfies properties (i) to (iv).
 
To check the first inequality in \eqref{lem:lambda_m}, observe that, on one hand, Lemma \ref{lem:diam_jordan_arc} implies $\diam(\Delta)\leq2R_{\Tt^l}\Ll(\Gamma,\Tt^l)$. On the other hand, since $\Delta$ contains a tile of $\Tt^{m-1}$, it follows that $2r_{\Tt^m}\leq \diam(\Delta)$. The inequality is now obtained by replacing $R_{\Tt^l}=\lambda^l R_{\Tt}$ and $r_{\Tt^{m-1}} = \lambda^{m-1}r_{\Tt}$ in the previous inequalities.

It remains to check the second inequality in \eqref{lem:lambda_m}. From the construction of the $P_l$'s, it is clear that each tile in $P_l$ is contained in a tile  of $\Tt^{l+1}$ that meets $\Gamma$ (which is the boundary of $\Delta$). Since each tile of $\Tt^{l+1}$ is subdivided in tiles of $\Tt^l$ according to the substitution rule, it follows that 
\[\Nn(\Delta_l,\Tt^l) \leq\, \|M\|_1 \Ll(\Gamma,\Tt^{l+1}).\]
\end{proof}

\begin{figure}[ht]
	\centering
	\begin{center}
		\begin{tabular}{cc}
			\includegraphics[scale=0.2]{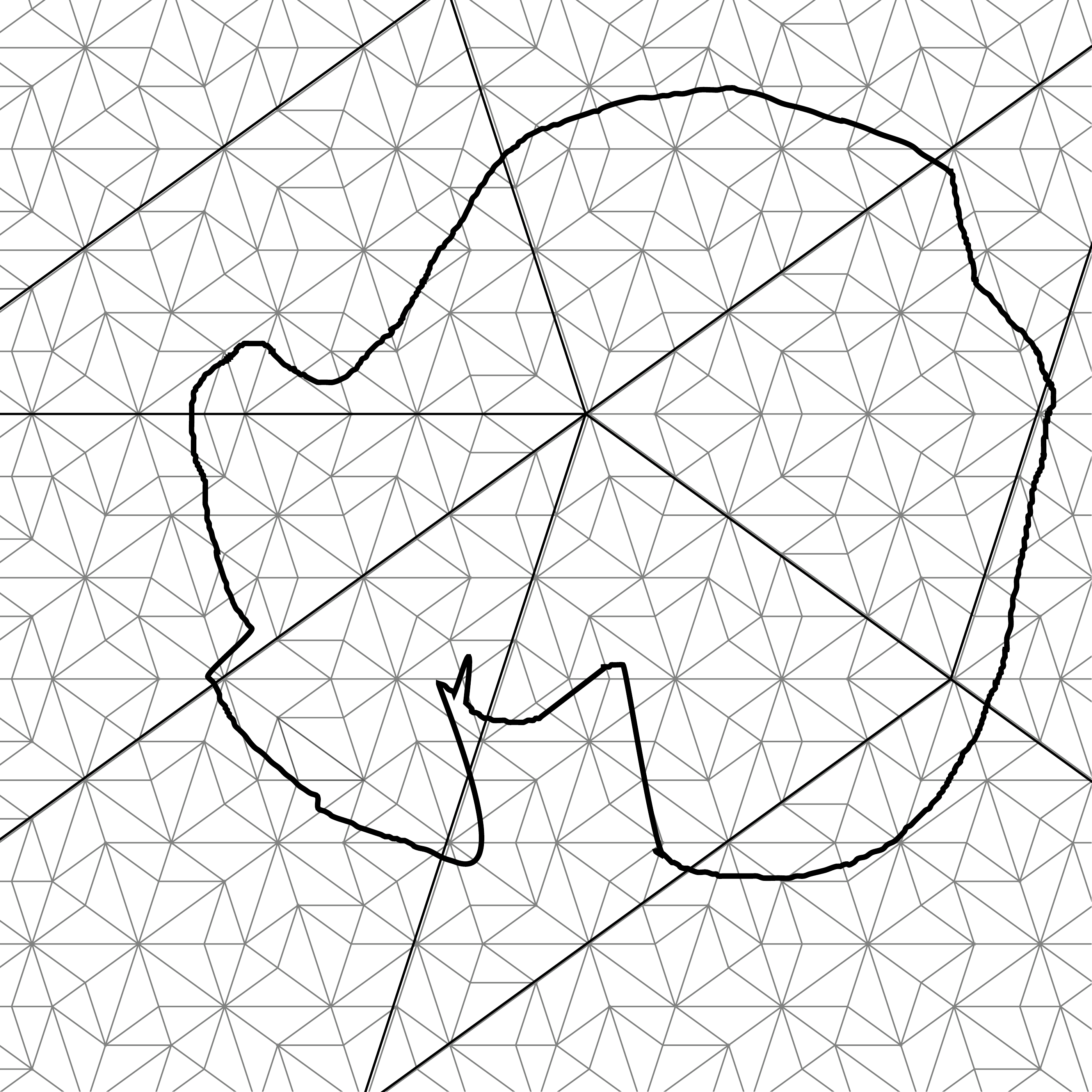} &
			\includegraphics[scale=0.2]{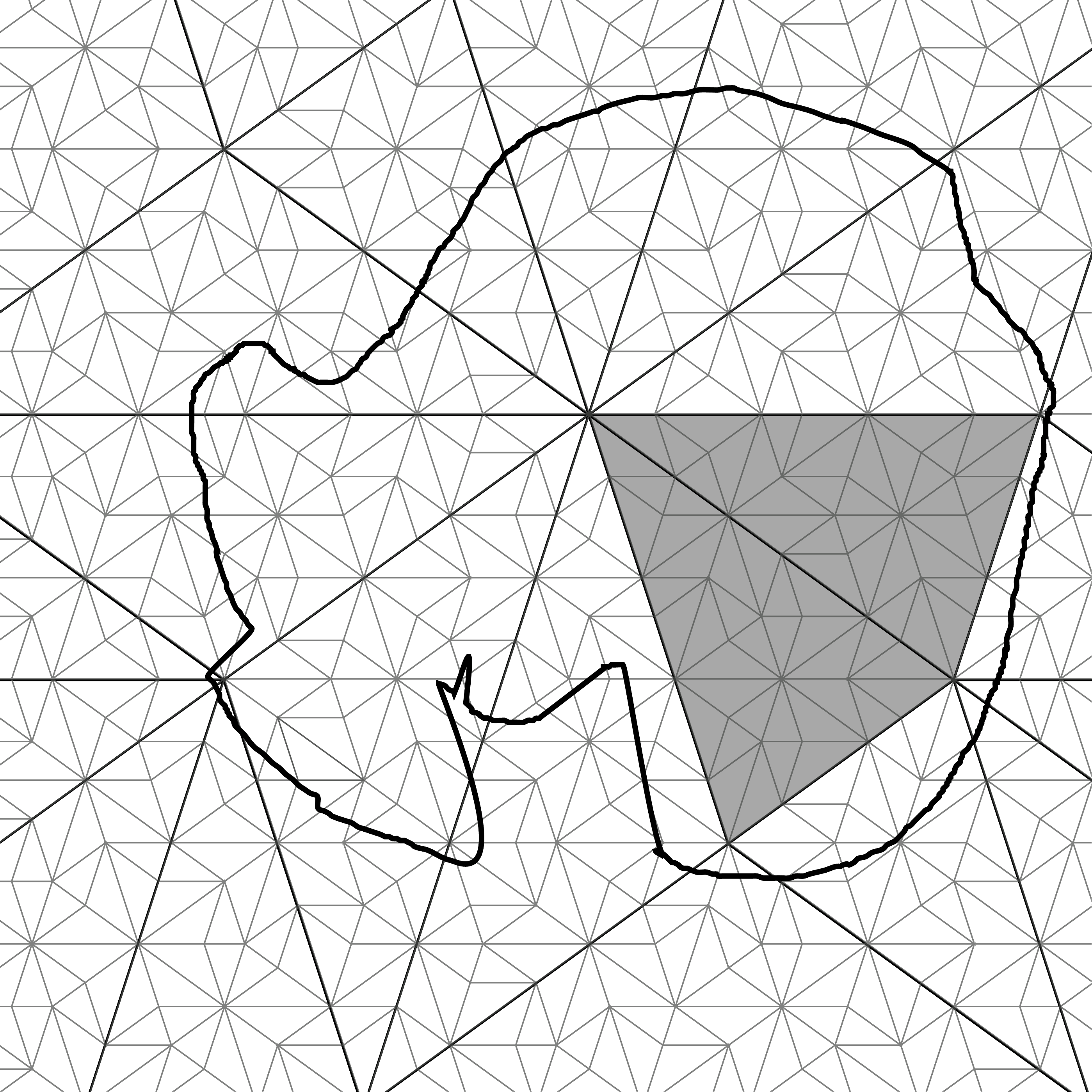} \\
			a) $l = m$&b) $l = m-1$\\
			\includegraphics[scale=0.2]{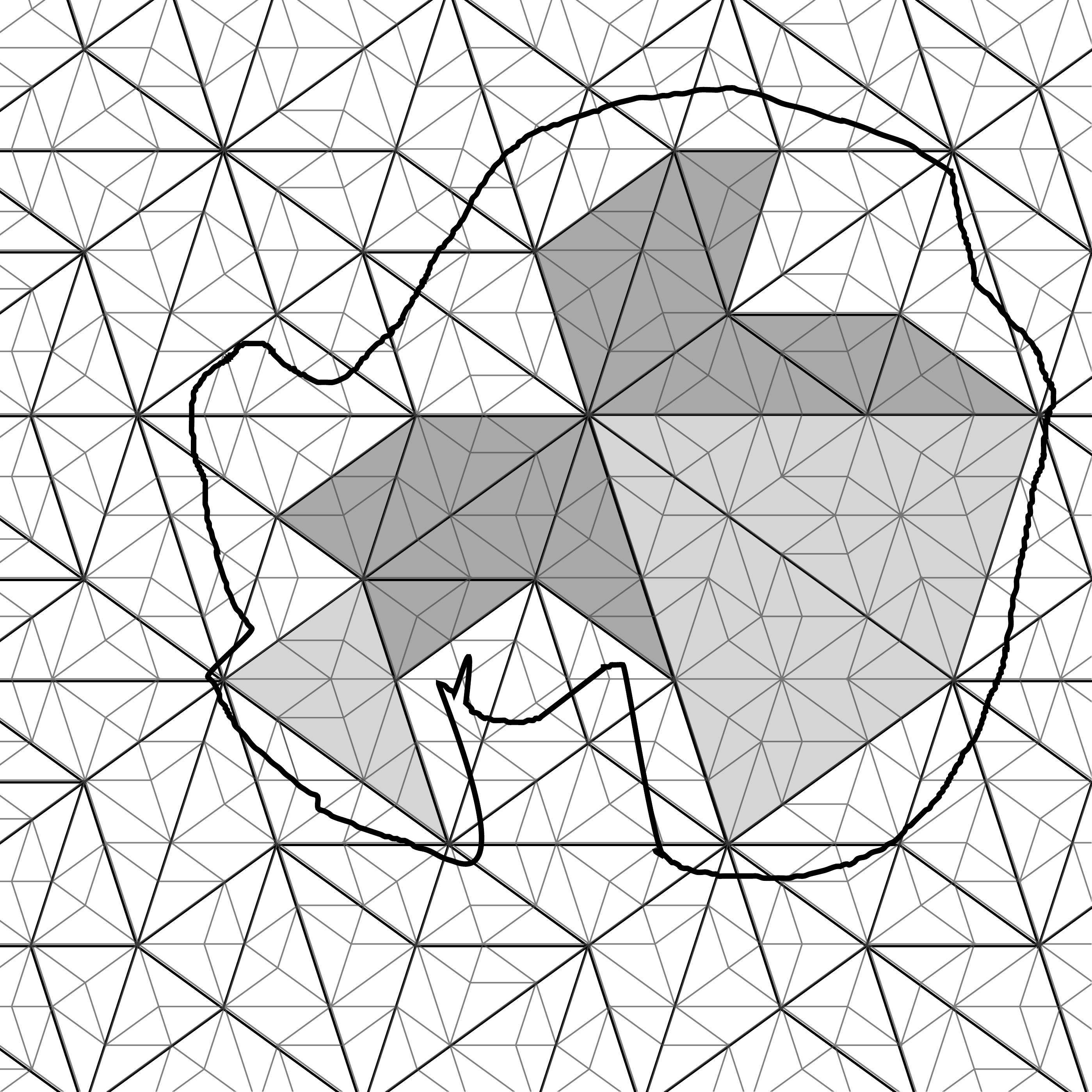}  &
			
			\includegraphics[scale=0.2]{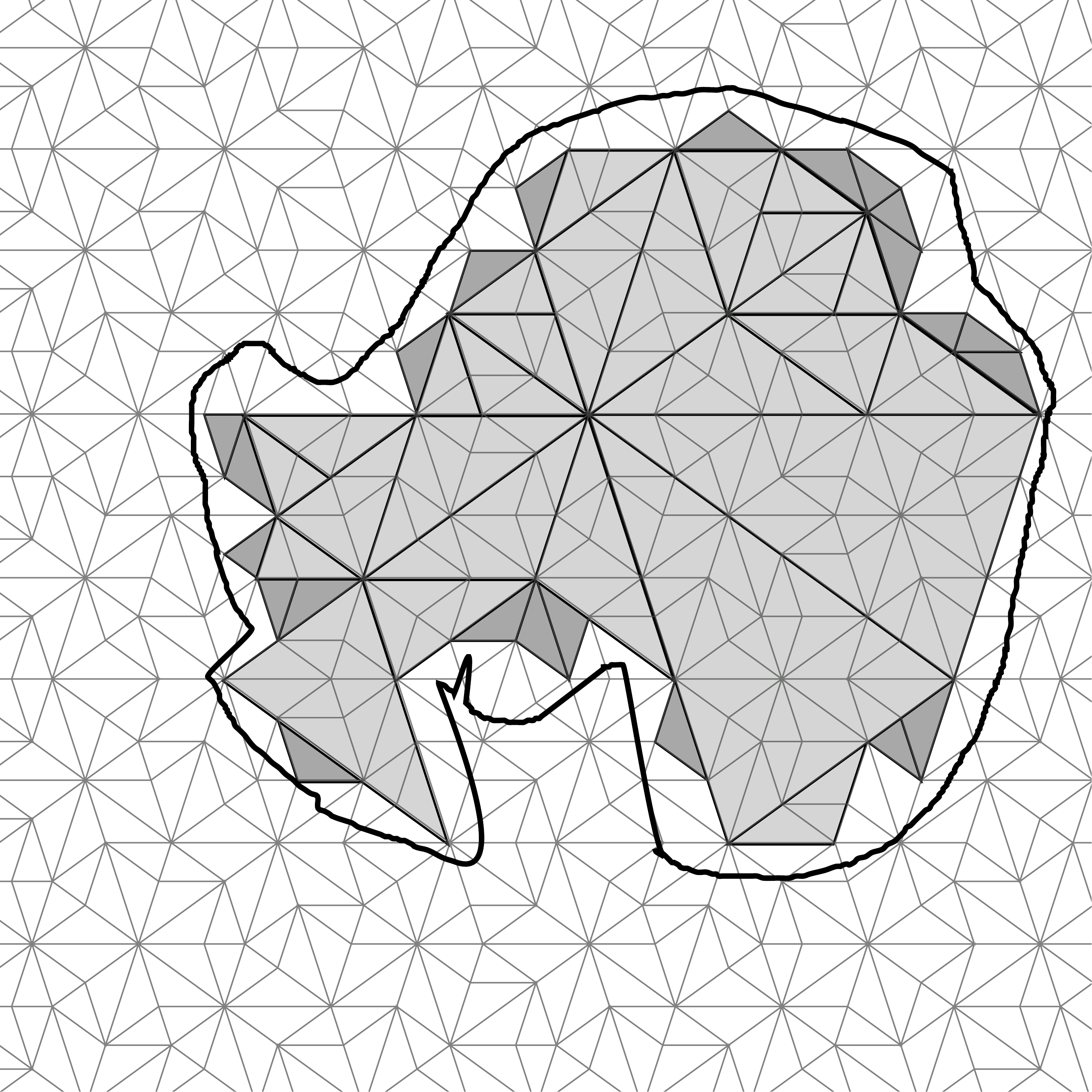}\\
			c)  $l = m-3$ &d) $l=0$
		\end{tabular}
		\label{tab:}
	\end{center}

	\caption{Construction of the hierarchical sequence for a given $\Delta$. The tiling in gray $\Tt^0$ is a Penrose tiling (see Section \ref{examples}). The tiling in black is $\Tt^l$, the dark-gray region is $\Delta_l$ and the light-gray one is $\widehat{\Delta}_{l+1}$. In this example, $m = 5$.}
\end{figure}

\subsection{Perron-Frobenius Theory}
\label{sec:PF}
We now recall the basic Perron-Frobenius theory that we will need in the sequel. For proofs see \cite[Chapter 8]{HornJohnson}. Let $M = M_\Ss$ be the substitution matrix associated with $\Ss$. Since $\Ss$ is primitive, the matrix $M$ is a primitive matrix with non-negative integer coefficients. The Perron-Frobenius Theorem  states that the largest real eigenvalue $\mu > 0$ of $M$, which is called the \emph{Perron eigenvalue}, is simple and greater than one. Moreover, there exists a right eigenvector $v^T = (v_1,\ldots, v_n)$ and a left eigenvector $w = (w_1,\ldots,w_n)$ such that $v$ and $w$ have positive coefficients and $\langle v, w\rangle = 1$. We denote by $r(M)$  the modulus of the second largest eigenvalue of $M$,  that is, 
\[r(M) = \max \{|\eta|\mid \eta \neq \mu\text{ is an eigenvalue of } M \}.\]

Recall that an eigenvalue is called \emph{semi-simple}  if its algebraic multiplicity is equal to its geometric multiplicity. 
The following proposition, or more precisely the corollary below, will be crucial for the proof of our main result.
The first part is a well-known consequence of the Perron-Frobenius Theorem, see for instance \cite[Theorem 8.5.1]{HornJohnson}. The second part is less well-known, but can be easily deduced from the first part by using the Jordan Canonical Form of $M$ after recalling that the Jordan blocks 
associated with semi-simple eigenvalues are diagonal. For $l>0$ and $i,j\in\{1,\ldots,n\}$, $m_{i,j}^l$ denotes the $(i,j)$-element of the matrix $M^l$. 

\begin{prop}\label{PF}
For every  $\rho > r(M)$ there exists $K = K(\rho)>0$ such that
\begin{equation}
\label{eq.pf1}
\vert m_{i, j}^l- v_i \mu^l w_j \vert \leq K \rho^l, 
\end{equation}
for all $l>0$. Moreover, if the eigenvalues of modulus $r(M)$ are semi-simple, then there exists $K>0$ such that 
\eqref{eq.pf1} holds with $\rho = r(M)$. 
\end{prop}
\begin{cor}\label{algebra}
Let $\nu_i = v_i/\sum_j v_j$ for all $i\in\{1,\ldots,n\}$. Then, for all $\rho > r(M)$, there exists a constant $K>0$ such that for all $i\in \{1, \dots, n\},$
\begin{equation}
\label{eq.pf2}
\vert m_{i, j}^l-\nu_i\sum_{k=1}^{n} m_{k, j}^l \vert \leq K \rho^l, 
\end{equation}
for all $l>0$. Moreover, if all the eigenvalues with modulus $ r(M) $ are semi-simple, then 
there exists a constant $K>0$ such that \eqref{eq.pf2} holds with $\rho = r(M)$.
\end{cor}
\begin{proof}
An easy computation shows
\[
\vert m^l_{i,j} - \nu_i  \sum_{k=1}^{n} m_{k, j}^l 
\leq     \vert m^l_{i,j} - v_i  \mu^l w_j  \vert +  
\frac{v_i}{\sum_j v_j}    \sum_{k=1}^n \vert v_k \mu^l w_j -  m_{k, j}^l \vert.
\]
The conclusion now follows from applying Proposition \ref{PF} twice.
\end{proof}

\section{Proof of the main result}
Let  $\Pp =\{p_1, \dots, p_n\}$ be  a finite set of tiles and  $\Ss$ a primitive substitution rule with dilation factor $\lambda>1$. Suppose that $\Tt$ is an admissible tiling for $\Ss$ and $\Gamma$ is a Jordan curve in $\mr^2$ bounding a closed disk $\Delta$.  

The idea of the proof is as follows. Fix $i\in\{1,\ldots,n\}$ and let $\nu_i$ be defined as in Corollary \ref{algebra}. First, it is not difficult to check that the number of $\Ee$-copies of $t_j$ after applying the substitution $l$ times to $t_i$ is exactly $m_{i,j}^l$. It follows that the density of the tile $t_i$ is $\nu_i$. Second, we consider a hierarchical sequence $(\Tt^l=(t^l_j)_{j\ge 0})_{l\geq 0}$ of $\Tt$ and the hierarchical decomposition $\{\Delta_0, \dots , \Delta_{m-1}\}$ of $\Delta$ constructed in Proposition \ref{prop.hier}. We use the decomposition of $\Delta$ to estimate $\Nn(\Delta,\Tt,p_i)$ and $\Nn(\Delta,\Tt)$. Since the sets $\Delta_l$ have disjoint interiors, we have 
\begin{equation}
\label{eq:syn_1}
\Nn(\Delta, \Tt, p_i) = \sum_{l=0}^{m-1}\Nn(\Delta_l, \Tt, p_i) = \sum_{l=0}^{m-1}\sum_{t_j^l\subseteq \Delta_l}\Nn(t_j^l, \Tt, p_i)
\end{equation}
and
\begin{equation}
\label{eq:syn_2}
\Nn(\Delta, \Tt) =\sum_{l=0}^{m-1}\sum_{t_j^l\subseteq \Delta_l}\Nn(t_j^l, \Tt).
\end{equation}
Next, multiplying \eqref{eq:syn_2} by $\nu_i$ and then subtracting \eqref{eq:syn_1}, we get 
\begin{equation}
\label{eq:syn_2_1} \Nn(\Delta, \Tt, p_i) - \nu_i (\Nn(\Delta, \Tt) )   = \sum_{l=0}^{m-1}\sum_{t_j^l\subseteq \Delta_l}\Nn(t_j^l, \Tt, p_i)
 - \nu_i \Nn(t_j^l, \Tt).
\end{equation}
Suppose that $t_j^l$ is a $\Ee$-copy of $\lambda^lp_k$ for some $k\in\{1,\ldots,n\}$. Then $\Nn(t_j^l, \Tt,p_i) = m_{i,k}^l$ and  
$\Nn(t_j^l, \Tt) = \sum_i m_{i,k} ^l$. Therefore, applying Corollary \ref{algebra} to \eqref{eq:syn_2_1}, we get 
\[\vert \Nn(\Delta, \Tt, p_i) - \nu_i  \Nn(\Delta, \Tt)\vert\,\leq\,  \sum_{l=0}^{m-1}\sum_{t_j^l\subseteq \Delta_l} K \rho^l\,\leq\,   \sum_{l=0}^{m-1}\Nn(\Delta_l,\Tt^l)K \rho^l,\]
where $K>0$, and either $\lambda > \rho > r(M)$ if $r(M)<\lambda$ or $\rho = \lambda$ if $r(M) = \lambda$ and the eigenvalues of modulus $r(M)$ are semi-simple. Applying Proposition \ref{prop.hier} to the last inequality, we obtain:
\begin{equation}
\label{eq:syn_4}\vert \Nn(\Delta, \Tt, p_i) - \nu_i  \Nn(\Delta, \Tt)\vert\leq   K\|M\|_1 \sum_{l=0}^{m-1}\Ll(\Gamma,\Tt^{l+1}) \rho^l.
\end{equation}

We want to apply Lemma \ref{lem:bord} to give an upper bound of $\Ll(\Gamma,\Tt^l)$ in terms of $\Ll(\Gamma,\Tt)$ for all $l\in\{0,\ldots,m\}$. Since for each $l\geq 0$, the tiles of $\Tt^{l+1}$ 
are tiled by tiles of $\Tt^l$, it follows that $\Ll(\Gamma,\Tt^{m-1})\leq \ldots\leq\Ll(\Gamma,\Tt^1)\leq\Ll(\Gamma,\Tt)$. It is easy to check that $K_{\Tt^l}=K_\Tt$ for all $l\geq 0$. 

Next, we define $l_0\in\{0,\ldots,m\}$ as follows. If  $\Ll(\Gamma,\Tt^{m-1}) > K_{\Tt}$, then $\ell_0:=m$. Else, we define $l_0$ to be the minimal $l\in\{0,\ldots,m-1\}$ such that $\Ll(\Gamma,\Tt^{l_0}) \leq K_\Tt$. Observe that in both cases $\Ll(\Gamma,\Tt^{l_0-1}) > K_\Tt$ for all $l \in\{0\ldots,l_0-1\}$.
  
We split the sum in \eqref{eq:syn_4} into two parts
\begin{equation}
\label{eq:syn_split}
\sum_{l=0}^{l_0-1}\Ll(\Gamma,\Tt^l) \rho^l +  \sum_{l=l_0}^{m-1}\Ll(\Gamma,\Tt^l) \rho^l
\end{equation}
and deal with each part separately. For the first sum, applying Lemma \ref{lem:bord}, we get
\[\Ll(\Gamma,\Tt^l)\leq (2K_{\Tt^l}+1)\frac{R_\Tt}{R_{\Tt^l}}\Ll(\Gamma,\Tt^l)\]
for all $l\in\{0,\ldots,l_0-1\}$. Since $R_\Tt/R_{\Tt^l}=\lambda^{-l}$ for every $l\geq 0$, it follows that
\begin{equation}
\label{eq:syn_first_sum}
\sum_{l=0}^{l_0-1}\Ll(\Gamma,\Tt^l) \rho^l \leq  (2K_\Tt+1)\Ll(\Gamma,\Tt) \sum_{l=0}^{l_0-1}\left(\frac{\rho}{\lambda}\right)^l. 
\end{equation}

To estimate the second sum in \eqref{eq:syn_split}, we suppose that $l_0 < m$ (otherwise, the sum is zero). Since $\rho\leq\lambda$ and $\Ll(\Gamma,\Tt^l)$
is decreasing, we have
\begin{equation*}
\sum_{l=l_0}^{m-1}\Ll(\Gamma,\Tt^l) \rho^l
\leq\Ll(\Gamma,\Tt^{l_0})\sum_{l=l_0}^{m-1} \lambda^l = \Ll(\Gamma,\Tt^{l_0})
\left(\frac{\lambda^m-\lambda^{l_0}}{\lambda-1}\right).
\end{equation*}
From Lemma \ref{lem:lambda_m} and the fact that $\Ll(\Gamma,\Tt_{l_0})<K_T$, we get $\lambda^{m-1-l_0}\leq K_\Tt R_\Tt/r_\Tt$. Denote $N_\Tt:= (\lambda K_\Tt R_\Tt/r_\Tt-1)/(\lambda-1)$. It follows that 
\begin{equation}
\label{eq:syn_6}
\sum_{l=l_0}^{m-1}\Ll(\Gamma,\Tt^l) \rho^l\leq N_\Tt\lambda^{l_0} \Ll(\Gamma,\Tt^{l_0}), 
\end{equation}
If $l_0>0$, then from Lemma \ref{lem:bord} and \eqref{eq:syn_6}, we get
\begin{equation}
\label{eq:syn_second_sum}
\sum_{l=l_0}^{m-1}\Ll(\Gamma,\Tt^l) \rho^l\leq (2K_\Tt+1)
N_\Tt\Ll(\Gamma,\Tt).
\end{equation}
It is clear that \eqref{eq:syn_second_sum} also holds when $l_0=0$, since  $(2K_\Tt+1)>1$. Combining 
\eqref{eq:syn_first_sum} and \eqref{eq:syn_second_sum}, we get
\begin{equation}
\label{eq:syn_third}
\vert \Nn(\Delta, \Tt, p_i) - \nu_i  \Nn(\Delta, \Tt)\vert\leq
 (2K_\Tt+1)\Ll(\Gamma,\Tt) \left(\sum_{l=0}^{l_0-1}\left(\frac{\rho}{\lambda}\right)^l + 
N_\Tt\right).\end{equation}
Finally, we deal with the  different cases for $\rho$. In the first case, we have $r(M_\Ss) <\rho < \lambda$ and it follows from \eqref{eq:syn_third} that 
\[
\vert \Nn(\Delta, \Tt, p_i) - \nu_i  \Nn(\Delta, \Tt)\vert\leq (2K_\Tt+1)\left(\frac{\lambda}{\lambda-\rho} + 
N_\Tt\right)\Ll(\Gamma,\Tt),
\]
which finishes part (i) of the main result, since $R_\Tt,r_\Tt$ and $K_\Tt$ are constant in $\Omega_\Ss$. 
In the second case, we have $\rho = \lambda$ and then from \eqref{eq:syn_third} and the estimation of $m\geq\ell_0$ given in Proposition \ref{prop.hier}, we obtain

\begin{equation*}
\vert \Nn(\Delta, \Tt, p_i) - \nu_i  \Nn(\Delta, \Tt)\vert\leq
 (2K_\Tt+1)\Ll(\Gamma,\Tt) \left( \log_\lambda\Ll(\Gamma,\Tt)  + \tilde{N}_\Tt\right),
\end{equation*}
where $\tilde{N}_\Tt = N_\Tt + 1 + \log_\lambda(R_\Tt/r_\Tt)$. 
Thus, there exists $K>0$ such that

\[
\vert \Nn(\Delta, \Tt, p_i) - \nu_i  \Nn(\Delta, \Tt)\vert\leq
K\Ll(\Gamma,\Tt)\ln\Ll(\Gamma,\Tt),  \]
  as soon as $\Gamma$  meets at least 2 tiles. This completes the proof of part (ii) of the main result. 
\section{Examples}\label{examples}
\subsection{Penrose tilings}
\begin{figure}
\begin{center}
	\includegraphics[scale=0.6]{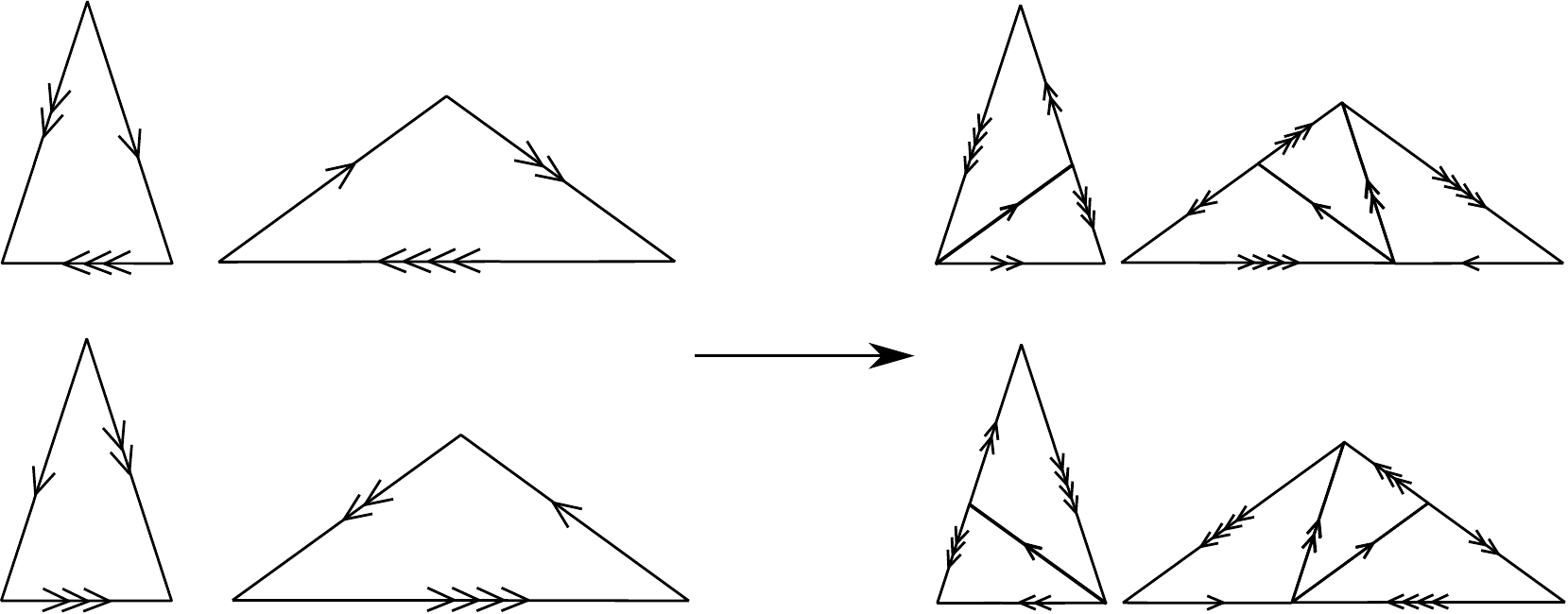}
	\captionof{figure}{Local and substitution rules to construct Penrose tilings}
	\label{fig:penrose}
\end{center}
\end{figure}

Penrose tilings are among the most known examples of self-similar tilings, see for instance \cite{Senechal,GS,Penrose,AP}. Here, we follow the construction of Anderson and Putnam \cite{AP}. Consider two isosceles triangles $t_1$ and $t_2$ in $\mr^2$,
where the vertices of $t_1$ have coordinates $(\sin({\pi}/{10}), 0), (-\sin({\pi}/{10}), 0)$ and $ (0, \cos({\pi}/{10}))$
and those of $t_2$ have coordinates $(\sin({3\pi}/{10}), 0), (-\sin({3\pi}/{10}), 0)$ and $ (0, \cos({3\pi}/{10}))$.
Both triangles are equipped with decorations (arrows) on their edges as shown in Figure \ref{fig:penrose}.

Let $\Ee$ denote the group of translations in $\mr^2$ and $\Pp_\text{Penrose}$ be the set of $40$ triangles obtained by rotation  of one of the two triangles $t_1$ or $t_2$ with its prescribed 
decoration by an angle $k\pi/10$, where $k$ is in $\{0,1, \dots , n-1\},$ and let $\Omega_\text{Penrose}$ 
be the set of tilings of $\mr^2$ made with translated copies of triangles in $\Pp_\text{Penrose}$ such that their tiles meet full-edge to full-edge and the decorations on overlapping edges coincide.  Elements of $\Omega_\text{Penrose}$ are called Penrose tilings. They can also be constructed by using the substitution $\Ss_\text{Penrose}$ described in Figure \ref{fig:penrose}. The dilation factor of $\Ss_\text{Penrose}$ is the golden mean $\lambda = (1 +\sqrt {5})/{2}$. 

The substitution matrix $M_\text{Penrose}$ is a $40\times 40$ non negative primitive matrix (see \cite{AP} for details). It is well-known that the Perron eigenvalue of $M_\text{Penrose}$ is equal to $\lambda^2  = (3 +\sqrt {5})/{2}$, and that $\lambda$ and $ -\lambda$ are the only eigenvalues of modulus $r(M_\text{Penrose})$. Straight-forward computations show that both of them are semi-simple with multiplicity $3$. Thus, we can  apply Theorem \ref{main} (part (ii)), and obtain:

Each triangle $p_i$ in $\Pp_\text{Penrose}$ has a well-defined frequency $\nu_i>0$, and there exists $K >0$ such that, for every Jordan curve $\Gamma$   and every  tiling $\Tt$ in $\Omega_\text{Penrose}$, we have:
\[\vert \Nn(\Delta,\Tt,  p_i) \, -\, \nu_i\Nn(\Delta,\Tt)\vert\, \leq\, K\Ll(\Gamma,\Tt)\ln(\Ll(\Gamma,\Tt),\]
where $\Delta$ is the closed disk bounded by $\Gamma$.

Now, Let  $\Ee$ be the group of all isometries (direct and indirect) of $\mr^2$. Up to these isometries, there are now only 2 types of Penrose tiles (the fat and thin triangles). The associated substitution matrix reads
\[ \begin{pmatrix}2 & 1  \\
 1 & 1\end{pmatrix},\]
and $({3+\sqrt 5})/{2}$ and $({3-\sqrt 5})/{2}$ are its eigenvalues. Thus, we can apply Theorem \ref{main} (part (i)) to obtain: 
\[\vert \Nn(\Delta,\Tt,  p_i) \, -\, \nu_i\Nn(\Delta,\Tt)\vert\, \leq\, K \Ll(\Gamma,\Tt).\]
\subsection{The square and table tilings} \label{square}
\begin{figure}
\includegraphics{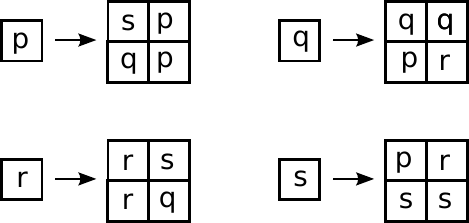}
\caption{Square table substitution}
\label{fig:square}
\end{figure}

Squares and table tilings were studied by Robinson \cite{Rob,rob99table}. First, we consider the square tilings. 
Let $\Pp_\text{square}$ be the set of four unit squares respectively decorated with symbols $p, q, r$ and $s$. The square substitution rule 
$\Ss_\text{square}$ is described in Figure \ref{fig:square}. The dilation factor is $\lambda = 2$.
In this case, the substitution matrix $M_\text{square}$ reads   
\[\begin{pmatrix}2 & 1 & 0 & 1 \\
1 & 2 & 1 & 0 \\
0 & 1 & 2 & 1 \\
1 & 0 & 1 & 2\end{pmatrix}.\]
The eigenvalues of $M_\text{square}$ are $\lambda^2$, $\lambda$ and $0$. It is easy to check that $\lambda$ is semi-simple with multiplicity $2$. Applying Theorem \ref{main} (part (ii)) (extended to deal with the decorated case), it follows that each square $t \in \Pp_\text{square}$  has a well-defined frequency  $\nu_t>0$ and there exists $K>0$ such that for any Jordan curve $\Gamma$  bounding a close disk $\Delta$ and any tiling $T$ in $\Omega_{\Ss_\text{square}}$, we have:
\[\vert \Nn(\Delta,\Tt,  t) \, -\, \nu_t\Nn(\Delta,\Tt)\vert\, \leq\, K \Ll(\Gamma,\Tt)\ln(\Ll(\Gamma,\Tt)).\]

\begin{figure}
\begin{center}
	\includegraphics{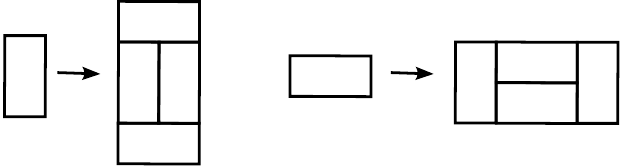}
	\captionof{figure}{The table substitution}
	\label{fig:table}
\end{center}
\end{figure}

Now we consider the table tilings. Let $\Pp_\text{table}=\{p_1, p_2\}$ where $p_1$ is a vertical domino and $p_2$ is a horizontal domino. The substitution rule $\Ss_\text{table}$ is described in Figure \ref{fig:table}. In this case, the dilation factor $\lambda$ is $2$ and the associated substitution matrix reads \[\begin{pmatrix}
   2 & 2 \\
   2 & 2 
\end{pmatrix}.\]
Its eigenvalues are $\lambda^2$ and $0$. Applying Theorem \ref{main} (part (i)), it follows that
each domino $p_i$ has a well-defined frequency $\nu_i>0$, and there exists $K>0$ such that, for any Jordan curve $\Gamma$  bounding a close disk $\Delta$ and any tiling $T$ in $\Omega_{\Ss_\text{table}}$, we have:
\[  \vert \Nn(\Delta,\Tt,  p_i) \, -\, \nu_i\Nn(\Delta,\Tt)\vert\, \leq\, K \Ll(\Gamma,\Tt).\]

\begin{figure}
\includegraphics{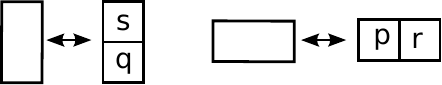}
\caption{Identification of table and square tilings}
\label{fig:mld}
\end{figure}

The square and table tilings can be identified (see \cite{Rob} for details) by identifying tiles 
as described in Figure \ref{fig:mld}.

This identification can be used to count tiles in the different tilings. Consider for instance the unit square marked with a $p$. In every tiling in $\Omega_{\Ss_\text{square}}$, the squares marked with $p$ and $r$ always appear as in Figure \ref{fig:mld}. Thus, for any closed Jordan curve $\Gamma$  bounding a close disk $\Delta$ we have: $\Nn(\Delta,\Tt,  p) - \Nn(\Delta,\Tt,  p_1)$ is smaller than $ \Ll(\Gamma,\Tt)$, where $p_1$ is the horizontal domino in $\Pp_\text{table}$. Hence, using the estimates for the table substitution gives better estimates than the estimates for the square substitution.  It is natural to wonder why? The answer simply relies on the fact that the square substitution system is less efficient than the table one in the sense that the estimates of the error for the $n$th substitution of the square $p$, which is a square of size $2^n$,  yields an error term of size $\lambda^n= 2^n$ . This error term is actually going to 
be canceled by terms corresponding to the other squares of size $2^n$. In our computation, the error term for a collection of squares of size $2^n$ is bounded from above by the sum of the error terms for each square of size $2^n$ and thus we cannot see these cancellations which actually occur. 
\subsection{The pinwheel tilings}
\begin{figure}
\begin{center}
	\includegraphics[scale=0.4]{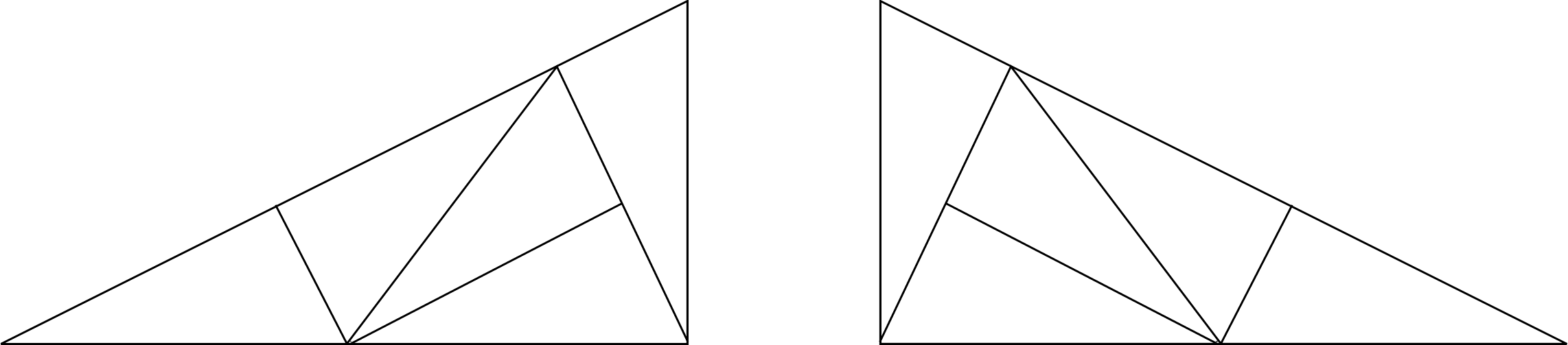}
	\captionof{figure}{The pinwheel substitution}
	\label{fig:pinwheel}
\end{center}
\end{figure}

Pinwheel tilings were introduced by  Conway and Radin  \cite{Radin}. Let $\Ee$ be the group $\Ii^+$ of direct isometries of $\mr^2$, and  $\Pp_\text{pinwheel}$ be the set of two right triangles which are isometric (but with inverse orientation). Each triangle has a hypotenuse oflength $1$ and sharp angle $\alpha$  satisfying $\tan \alpha  = {1}/{2}$. The substitution rule $\Ss_\text{pinwheel}$ is described in Figure \ref{fig:pinwheel} and has a dilation factor of $\sqrt 5$.

The substitution matrix is the  $2\times 2$ matrix $\begin{pmatrix}
   2 &3 \\
  3 & 2 
\end{pmatrix}$ 
It has two eigenvalues $\lambda^2 =5$ and $-1$. Applying Theorem \ref{main}, it follows that each triangle  $t_i$ in $\Pp_\text{pinwheel}$ has a well-defined frequency $\nu_i>0$, and 
there exists $K >0$  such that, for any Jordan curve $\Gamma$  bounding a close disk $\Delta$ and any tiling $T$ in $\Omega_{\Ss_\text{pinwheel}}$:
\[  \vert \Nn(\Delta,\Tt,  p_i) \, -\, \nu_i\Nn(\Delta,\Tt)\vert\, \leq\, K \Ll(\Gamma,\Tt).\]

\subsection{Rauzy tilings}
Rauzy tilings (see \cite{Rau} for details) are tilings made from a set $\Pp_\text{rauzy}$ of three (topological) disks $r_1, r_2$ and $r_3$, where $r_1$ is a disk whose boundary is a Jordan curve (with positive Hausdorff dimension), $r_2 = \lambda r_1$ and $r_2 = \lambda^2 r_3$ where $\lambda\simeq 1.84  $ is the unique real root of the polynomial $X^3 -X^2-X-1$ (see \cite{Pit}).  
The substitution matrix is the  $3\times 3$ matrix 
\[\begin{pmatrix}
 0 &0&1\\
 1&0&0\\0&1&1 
\end{pmatrix}.\]
The largest eigenvalue of the substitution matrix is $\lambda^2 $. All other eigenvalues have modulus smaller than $1$. Applying Theorem \ref{main}, it follows that 
each Rauzy tile $p_i$ has a well-defined frequency $\nu_i$ and 
there exists $K > 0$ such that for every  Jordan curve $\Gamma$  bounding a close disk $\Delta$ and every  tiling $\Tt$ in $\Omega_{\Ss}$, 
we have:
\[  \vert \Nn(\Delta,\Tt,  p_i) \, -\, \nu_i\Nn(\Delta,\Tt)\vert\, \leq\, K\Ll(\Gamma,\Tt).\]
\noindent \textbf{Acknowledgements.} This work is part of the project {\it CrystalDyn} supported
by the "Agence Nationale de la Recherche" (ANR-06-BLAN- 0070-01). Part of this work was done while 
J. Aliste-Prieto was a Junior Research Fellow  at the Erwin Schr\"odinger Institute in Vienna. J. Aliste-Prieto also acknowledges funding from Fondecyt Postdoctoral Grant 3100097. D. Coronel is funded by Fondecyt Postdoctoral grant 3100092
and PBCT-Conicyt Research Project ADI-17. Finally, the authors  would like to thank Andrew Hart for his help in correcting the English in this article.


\end{document}